\documentclass[twocolumn,showpacs,preprintnumbers,amsmath,amssymb,superscriptaddress,nofootinbib]{revtex4}

\usepackage[english]{babel}
\usepackage{amsmath}
\usepackage{graphicx}
\usepackage{dcolumn}
\usepackage{bm}
\usepackage{amscd}
\usepackage{hyperref}
\usepackage[utf8x]{inputenc}
\usepackage{natbib}
\usepackage{color}

\usepackage{empheq} 
\usepackage{mathtools}

\begin{document}

\title{L\'evy flights with power-law absorption}

\author{Luca Cattivelli}
\email[]{luca.cattivelli@sns.it}
\affiliation{Scuola Normale Superiore, Pisa, Italy}

\author{Elena Agliari}
\email[]{agliari@mat.uniroma1.it}
\affiliation{Dipartimento di Matematica, Sapienza Universit\`a  di Roma}

\author{Fabio Sartori}
\email[]{fabio.sartori@brain.mpg.de}
\affiliation{Max Planck Institute for Brain Research, Frankfurt, Germany}

\author{Davide Cassi}
\email[]{davide.cassi@fis.unipr.it}
\affiliation{Dipartimento di Fisica, Universit\`a  di Parma}

\date{\today}

\begin{abstract}

We consider a particle performing a stochastic motion on a one-dimension lattice with jump lengths distributed according to a power-law with exponent $\mu + 1$. Assuming that the walker moves in the presence of a distribution $a(x)$ of targets (traps) depending on the spatial coordinate $x$, we study the probability that the walker will eventually find any target (will eventually be trapped). We focus on the case of power-law distributions $a(x) \sim x^{-\alpha}$ and we find that, as long as $\mu < \alpha$, there is a finite probability that the walker will never be trapped, no matter how long the process is. This result is shown via analytical arguments and numerical simulations which also evidence the emergence of slow searching (trapping) times in finite-size system. The extension of this finding to higher-dimension structures is also discussed.

\end{abstract}

\maketitle

\section{Introduction \label{introduction}}

Two-species reaction-diffusion processes provide a model for a number of physical, chemical, biological, and even social phenomena \cite{havlin}. In particular, biological encounters and random searches typically involve a mobile component and a reactive component, the latter being a prey, a mate, or a convenient place where settling \cite{Stanely-2011}. In this context, the kind of motion performed by the ``searcher'' plays a crucial role and, as evidenced by extensive experiments in the last two decades, anomalous diffusion seems to be widespread among real organisms \cite{Stanely-2011}. In fact, many animals, ranging from birds (e.g., albatross  \cite{albatross}), to arthropods (e.g., bees \cite{bees}, butterflies \cite{butterflies}), to aquatic animals (e.g., sharks, sea turtles, and penguins \cite{marine}), and even to mammals (e.g., deer \cite{albatross}, goats \cite{goat}) rarely display Gaussian probability functions for displacement with a variance scaling linearly with time, as prescribed by normal diffusion. Rather, they display L\'evy walk movement patterns \cite{Zumofen-CP1990,Blumen-EPL1990}. Also, many other natural systems beyond animals (e.g., pollen \cite{pollen} and seeds \cite{seed}), exhibit this kind of anomalous movement.

The prevalence of such a behaviour may be related to the fact that this turns out to be the optimal strategy for locating sparse resources in a wide range of environments \cite{Viswanathan-Nature1999,Reynolds-Ecology2009,Humphries-PNAS2012}. Notably, also humans have been shown to move according to L\'evy walks and this kind of superdiffusion plays an important role in pandemics \cite{Shlesinger-1995}. 

One usually distinguishes between L\'evy walks and L\'evy flights. Both are stochastic motion characterized by jump lengths distributed according to a power-law with exponent $\mu + 1$ [experimental studies have shown that these exponents are very close to $\mu=1$ (e.g., for snails $\mu$ is between $\mu=1.2$ and $\mu=1.7$ \cite{snail}, for deer $\mu=1.16$ \cite{deer} and for humans $\mu=0.75$ \cite{human})], yet in the former case the walker moves with a typical, or sometimes constant, velocity, namely the step length is proportional to the elapsed time, while in the latter case the time taken by any jump is independent of its length.
L\'evy walks are modeled by assuming a coupling between jump length and jump time (see e.g., \cite{West-PRL1987,Klafter-PRA1987}) and are regarded as the appropriate tool for describing superdiffusion \cite{godec2013finite,zimbardo2013levy}.

Here, we focus on L\'evy flights, as they can be more easily approached analytically, and we check numerically that the emerging features also hold for L\'evy walks.

In particular, we study encounter processes between a searcher (e.g., forager, predator, parasite, pollinator, or the active gender in the mating search), and a target (e.g., prey, food, or the passive gender in the mating activity) \cite{Zumofen-JSP1991}. The searcher is modeled as a particle performing a stochastic motion on a $d$-dimensional lattice, while the target(s) is assumed to be spread along the underlying space. The encounter occurs when the particle reaches any site occupied by a target \footnote{As remarked in \cite{Stanely-2011}, in statistical models of many biological processes the ``microscopic'' details are often essentially irrelevant to the averages and can be neglected. Of course, this implies some limitations about the applicability of such models.}. Otherwise stated, here the searcher and the targets do not interact along the jump but only at the end of this, namely at the (possible) turning point. In fact, in many contexts (e.g., as for pollination or fishing) this is a realistic hypothesis. 

Moreover, here we allow for non-uniform distributions of targets. More precisely, starting from a given position $r_0$, the searcher performs a L\'evy flight in a space where each site at a distance $r$ from $r_0$ is potentially occupied by a target with a probability $a(r)$. Otherwise stated, one can think that the particle is moving in the presence of dense traps whose absorbing rate is spatially inhomogeneous as ruled by $a(r)$. Thus, we wonder under which conditions, if any, there exists a finite probability that the particle is not able to reach any target, or, equivalently, that it will never be trapped. 

We especially focus on a searcher whose step lengths are distributed according to $p(r) \sim1/|r|^{1+\mu}$ in the presence of targets/traps effectively distributed according to a power-law $a(r)\sim 1/|r|^{\alpha}$. This means that the distribution of, say, food, potential mates, or predators follows a gradient or that the encounter has a rate of success which decreases with the distance from the origin.
The problem is first investigated for the one-dimensional substrate finding that, when $\mu \geq1$, namely when the searcher performs a recurrent diffusion process (i.e., it is  certain to visit all sites), the target is eventually found with probability equal to $1$. On the other hand, when $\mu <1$, namely when the searcher performs a transient diffusion process (i.e., it is not certain to visit all sites), \emph{and} $\alpha > \mu$ there is a finite probability that the searcher will never find the target. For higher-dimension lattices analogous regimes emerge. 
Numerical simulations also allow getting some insights for the case of finite structures and of truncated L\'{e}vy flights.

The paper is organized as follows: in Sec.~\ref{sec:LF} we provide a streamlined description of L\'evy flights and in Sec.~\ref{sec:trapping} we introduce a formulation of the problem in terms of a fractional diffusion equation. Then, in Secs.~\ref{sec:numerics} and \ref{sec:numerics2} we present our numerical simulations and the related results. A discussion about the general $d$-dimensional case can be found in Sec.~\ref{sec:higher}, while Sec. \ref{sec:peaks} contains further arguments corroborating our results. Sec.~\ref{sec:conclusions} is left for our conclusions and perspectives and, finally, Appendix \ref{sec:fractional}  contains some technical tools concerning fractional calculus.

\section{L\'evy flights} \label{sec:LF}
Let us consider a mobile agent on a line that, at each time step $t$, jumps in random direction to a distance $|x|$, taken from a power-law distribution:
\begin{equation} \label{eq:jumps}
p(x)\sim\frac{1}{|x|^{1+\mu}}.
\end{equation} 
It can be shown (e.g., see \cite{havlin}) that, for $\mu<2$ and for asymptotic times, the probability density $\rho(x,t)$ of the searcher being at $x$ at time $t$  follows the L\'evy distribution\footnote{When $\mu\geq2$ the motion displays (for asymptotic times) a Gaussian probability function with a variance scaling linearly with time, namely a normal diffusion.}, which for $|x|^\mu \gg t$ (considering dimensionless units) scales as:
\begin{equation} \label{eq:asymptotic}
\rho(x,t)\sim\frac{t}{|x|^{1+\mu}}.
\end{equation}
The resulting mean-square displacement $\langle x^2 \rangle \equiv \int_{-\infty}^{\infty} x^2 \rho(x,t) dx$ is therefore divergent. More generally, when embedded in a $d$-dimensional lattice, a mean-square displacement $\langle r^2 \rangle$ scaling with time faster than linearly can be accomplished by taking a distribution for step lengths scaling like $p(\vec{r}) \sim 1/r^{\mu +d}$, with $0<\mu <2$ (e.g., see \cite{brasilian}). As a result, the fractal dimension $d_f$ of the sites visited by this L\'evy flight on a line is $d_{f}= \mu$ for $ \mu < 1$ and $d_{f}=1$ for $\mu\geq1$ \cite{Levy-dimension}, and, in general, on $d$-dimensional lattices is $d_f = \mu$ for $\mu <d$ and $d_f = d$ for $\mu \geq d$ \cite{Hughes2,Fogedby}. Thus, when the mean displacement (averaged over the distribution $p(x)$) diverges algebraically ($\mu<1$) the agent does not visit every site of the line and the process is transient. A formal proof about transition between transience and recurrence depending on $\mu$ can be found in \cite{Hughes}. 

L\'evy flights can be effectively approached via the so-called fractional calculus which provides a generalization of classical diffusion equations using fractional derivatives, especially fractional Laplacian operators (see also \cite{Jespersen-PRE1999} and Appendix \ref{sec:fractional}).
More generally, we mention that the description of superdiffusion can be accomplished in terms of fractional material derivatives (see \cite{Metzler_Klafter, Sokolov-PRE2003}) which, as far as a dimensional analysis is carried out, give the same scalings as fractional derivatives.

\section{Power-law absorption probability on one-dimensional structures} \label{sec:trapping}

The L\'evy flight embedded in an infinite chain can be described by the following free-propagation equation
\begin{equation} \label{eq:free}
\frac{\partial\rho(x,t)}{\partial t}=\frac{\partial^\mu \rho(x,t)}{\partial |x|^\mu} ,
\end{equation}
where the information on the heavy-tailed distribution of jumps (see Eq.~\ref{eq:jumps}) is fully defined in the fractional derivative with respect to the variable $x$ \cite{Klafter-2011,Barkai-PRE2000,Barkai-ChemPhys2002,Barkai-PRE1997} encoded by the Riesz operator $\frac{\partial^\mu }{\partial |x|^\mu}$; the latter is defined in Appendix \ref{sec:fractional} and it recovers the standard $\mu$-th order derivative for integer $\mu$.

As anticipated, here we study a L\'evy flight in a inhomogeneous absorbing medium where at each site $x$ the walker has a probability $a(x)$ to be absorbed. Therefore, the equation for free propagation (\ref{eq:free}) is perturbed by adding a term proportional to $- \rho (x, t) a (x)$ (see e.g., \cite{BenNaim-JSP1993}), namely
\begin{equation} \label{eq:absorption_equation}
\frac{\partial\rho(x,t)}{\partial t}=\frac{\partial^\mu \rho(x,t)}{\partial |x|^\mu} -a(x)\rho(x,t).
\end{equation}

This system actually models different kinds of phenomena. For instance, we can have a space completely filled by partially absorbing traps whose absorbing rate depends on $x$ as $a(x)$, or we can have a space where targets are distributed in space according to a density $a(x)$. In any case the probability that, being the searcher on site $x$, the encounter (or the absorption) effectively occurs is $a(x)$. For this encounter (or absorbing) rate, here we adopt a general power-law distribution given by
\begin{equation} \label{eq:rate}
a(x)= \frac{K}{|x|^{\alpha}+1},
\end{equation} 
being $K$ a positive, real constant, hereafter set equal to $1$ without loss of generality. 

Now, the differential equation (\ref{eq:absorption_equation}) has no general analytical solution, but is is useful to discuss two limit cases for Eq.~\ref{eq:absorption_equation}, corresponding to a negligible diffusion term (e.g., the time-scale for diffusion is much longer than the time-scale for absorption) and to a negligible absorption term (e.g., the time-scale for absorption is much longer than the time-scale for diffusion).
\newline
In the former case $\partial\rho(x,t)/\partial t = - a(x) \rho(x,t) <0$, so that the distribution $\rho(x,t)$ tends to zero exponentially with time with rate $a(x)$. 
\newline
In the latter case, we recover the pure (fractional) diffusive equation (\ref{eq:free})
%
%
whose solution is known (see Eq.~\ref{eq:asymptotic}), and for large $x$ it is 
\begin{equation} \label{eq:normal}
\rho(x,t)\sim\frac{t}{|x|^{\mu+1}}.
\end{equation}

Further, we notice that, as mentioned in Sec.~\ref{sec:LF},  when $\mu \geq 1$, the motion is recurrent, that is, the particle visits \emph{any} site infinite times on average; this holds in particular also for sites belonging to the neighbourhood of the origin where $a(x)$ is finite and this ensures, eventually, the absorption. On the other hand, when $\mu <1$, the process is transient, that is, the number of returns to any site is finite and the particle spends most of its time on sites where $a(x)$ is vanishing. Consistently, when $\mu \geq 1$ the step distribution is less spread and we expect that the emerging motion is closer to the classical random walk. Indeed, a simple random walk on a one-dimensional lattice visits any site an infinite number of times and, in the presence of a distribution $a(x)$, is therefore eventually absorbed.

On can therefore wonder whether the encounter is ineluctable upon the introduction of the encounter rate (\ref{eq:rate}), whatever the value of $\alpha$ and $\mu$, or there exists a choice for these parameters such that survival can take place. This point is investigate numerically in the next section.


\section{ Numerical results for L\'{e}vy Flights} \label{sec:numerics}

We now check the asymptotic behaviour of the solution $\rho(x,t)$ of Eq.~\ref{eq:absorption_equation} for different values of $\alpha$ via numerical simulations, following two distinct routes. In the former we study the asymptotic probability of being eventually absorbed. In the latter, we analyze the asymptotic behaviour of the mean square distance covered and compare it with the asymptotic  behaviour expected without absorption. For both routes we simulate a L\'evy flights in a one dimensional space and whose step length $x$ is extracted from the probability distribution $p(x) =N(\mu)[|x|^{\mu+1}+1]^{-1}$, being $N(\mu)=[2\Gamma (\frac{\mu}{1+\mu})\Gamma (1+\frac{1}{1+\mu})]^{-1}$ the proper normalization factor.
After each jump the searcher has a probability $a(x)$  of being absorbed, where $x$ is its current position. The time $t$ is given by the number of jumps accomplished.

Let us start with the first route, where we explore the asymptotic behaviour of the absorption probability $F(\alpha, \mu)$ defined as 
\begin{eqnarray} \label{eq:F}
F(\alpha, \mu) =1-  \lim_{t \to +\infty}  \left ( \int_{-\infty}^{\infty} \mathrm{\, d}x \, \rho_{\alpha , \mu}(x,t) \right),
\end{eqnarray}
where we have highlighted the dependence on $\alpha$ and on $\mu$ of the distribution function $ \rho_{\alpha , \mu}(x,t)$.

\begin{figure}[b!]
\includegraphics[width=1.4\linewidth]{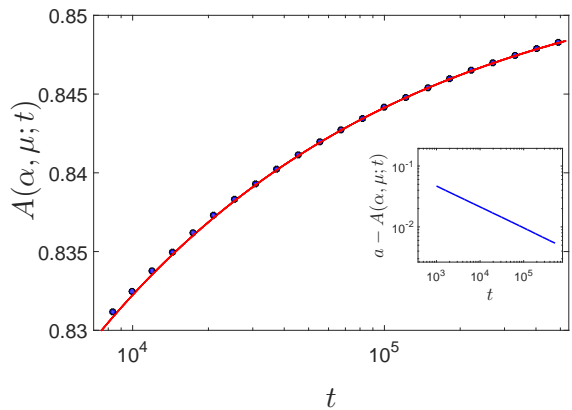}
\caption{(Color on line) Absorption probability $A(\alpha,\mu;t)$ for the case $\mu = 0.7$, $\alpha =1.1$. Data from numerical simulations (bullets) are fitted by the function $y = a-b t^{-c}$ (solid line). The best fit, corresponding to a goodness parameter $R^2 =0.99$ is given by the coefficient $a = 0.854 \pm 0.009$, $b=0.534 \pm 0.008$, and $c= 0.348 \pm 0.014$. In particular, $a$ provides an estimate for the asymptotic probability of being absorbed, that is $F(\alpha, \mu)$. Inset: We plotted the difference $a-A(\alpha,\mu;t)$, where $A(\alpha,\mu;t)$ is derived from numerical simulations and $a$ is the best-fit coefficient. In a log-log scale, this quantity depends linearly in time, corroborating the expected power law behaviour of $A(\alpha,\mu;t)$.}
 \label{frequency_absorption}
\end{figure}

The integral of Eq.~\ref{eq:F} is estimated with Monte Carlo simulations, as described hereafter. For every $\mu$ and $\alpha$ we run L\'evy flights and we measure the number of absorbed flights divided by the total number of simulated flights. An example of this quantity, referred to as $A(\alpha, \mu; t)$, is reported in Fig.~\ref{frequency_absorption}. This quantity is then fitted with a power law $y = a-b t^{-c}$, which is the typical time saturation law for this kind of processes \cite{havlin}. In fact, for all the cases analyzed, such a power law provides a successful description for the behaviour of $A(\alpha, \mu; t)$ with respect to time $t$.
The fit coefficient $a$, corresponding to the asymptotic value of $A(\alpha, \mu; t)$, just provides our estimate for $F(\alpha, \mu)$.

\begin{figure}[t!]
\includegraphics[width=1.4\linewidth]{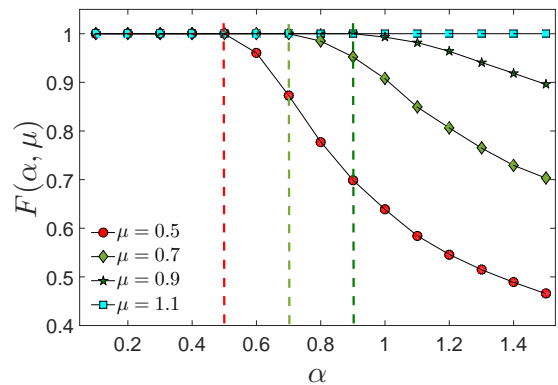}
\caption{(Color on line) The asymptotic absorption probability $F(\alpha,\mu)$ found from Monte Carlo simulations. For the cases analysed, the absorption is certain when $\mu=1.1$ and when $\alpha\leq \mu$ with $\mu=0.5$, $\mu=0.7$ and $\mu=0.9$. The solid line are guides for the eye. The dashed vertical lines correspond to $\alpha=0.5$, $\alpha=0.7$, and $\alpha=0.9$. The number of replicas is $10^7$ and the asymptotic regime is reached typically around $t_f=10^6$. The errors on $F(\alpha,\mu) $ are of order $10^{-4}$.
\label{F}}
\end{figure}

In Fig.~\ref{F} we show the asymptotic behaviour of the absorption probability $F(\alpha, \mu)$ versus $\alpha$. As long as $\alpha \leq \mu$, we get that $F(\alpha, \mu)=1$, meaning that  absorption is certain, while when $\alpha > \mu$, we get that $F(\alpha, \mu)<1$, meaning that there is a finite probability of surviving.
 
However, the condition $\alpha>\mu$ is not sufficient to avoid the certainty of the absorption: recalling what stated in the previous section, in addition one has to require that the process is not recurrent, namely that $\mu <1$. Summarizing, the particle can escape absorption or, analogously, has a finite probability of never finding the target as long as
%
%
%
\begin{empheq}[left=\empheqlbrace]{align}
&\alpha>\mu, \label{eq:condition}\\
&\mu <1. \ \label{eq:condition2}
\end{empheq}

\begin{figure} 
\includegraphics[width=1.4\linewidth]{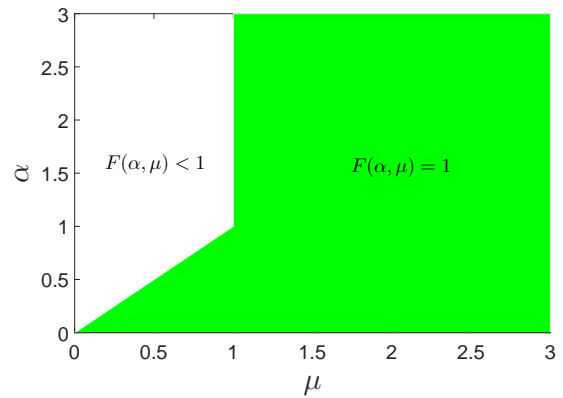}
\caption{(Color on line) Phase diagram on a line in the parameter space $(\alpha,\mu)$. The white region corresponds to a finite surviving probability, that is $F(\alpha, \mu) <1$, while the colored region corresponds to certain trapping, that is $F(\alpha, \mu) =1$.   }
\label{phase_diagram1}
\end{figure}

This result is depicted in a phase diagram, highlighting the regions of the plane $(\alpha,\mu)$ where the particle is either certain or not-certain to be trapped (see Fig.~\ref{phase_diagram1}).
A corroboration of this picture via analytical arguments can be found in Sec. \ref{sec:peaks}.

In the second route of numerical investigations  we analyze the asymptotic behaviour of the mean square displacement.
In  one dimension 
%
%
the asymptotic behavior of the distribution (without absorption) $\rho(x,t)$ is power law (see eq. \ref{eq:normal}) such that (for $\mu<2$) the mean square displacement $\langle x^2(t) \rangle$ diverges for any $t$.  
Therefore, $\langle x^2(t) \rangle$ is not a suitable quantity to characterize the scaling of the process. 
Alternatively, one could consider an imaginary growing box and measure the related mean square displacement $\langle x^2(t) \rangle_L$ given by  \cite{Metzler_Klafter}
\begin{equation}\label{eq:variance}
\left\langle x^{2}(t)\right\rangle _{L}\sim\int_{L_{1}t^{1/\mu}}^{L_{2}t^{1/\mu}}x^{2}\rho(x,t)\mathrm{\, d}x\sim t^{2/\mu},
\end{equation}
where $L_1$ and $L_2$ are proper finite constants. More precisely, the boundaries of the box (i.e., the extremes of the integral) are chosen in order to ensure that the asymptotic behavior $\rho(x,t)\sim t/ x^{\mu+1}$ (see Eq.~\ref{eq:normal}) is reached; indeed the asymptotic regime is valid only for  $|x|^\mu \gg t$. 
Moreover, for $ \mu> 1$, the squared absolute mean $\langle |x| \rangle^2 = (\int dx |x|\rho (x, t))^2$  is proportional to $\langle x^2 (t) \rangle_L$. 
%
For this reason the integral (\ref{eq:variance}) is also called pseudo or imaginary mean squared displacement \cite{Metzler_Klafter}.

Now, having an analytical estimate of $\left\langle x^{2}(t)\right\rangle _{L}$ in the absence of traps, we measure $\left\langle x^{2}(t)\right\rangle _{L}$ through numerical simulations accounting also for absorption.

%
%
The density $\rho(x,t)$ is numerically estimated by counting the number of realizations where the moving agent has survived and is in $x$ at time $t$, divided by the overall number of realizations (including those where the agent has been trapped or is outside the imaginary box).
Also for simulations, the values of $L_{1}$ and $L_{2}$ are chosen large enough so that the asymptotic regime is ensured and in order to produce good statistics for all values of $\alpha$ \footnote{We stress that here we say that the asymptotic regime is reached at time $t_f$ as long as the variation of $\rho(x,t)$ over a time interval spanning an order of magnitude, i.e., $(t_f, 10 t_f)$, is numerically negligeable.}. During the simulation, long jumps leading the process outside the box are not discarded, but the resulting density $\rho(x,t)$ will not contribute to the integral.

\begin{figure}
\includegraphics[width=1.4\linewidth]{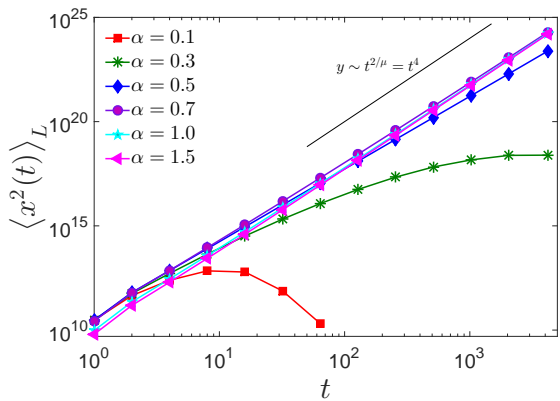}
\caption{(Color on line) Mean-square displacement $\left\langle x^{2}(t)\right\rangle _{L}$ obtained from Monte Carlo simulations performed on one-dimensional structures for fixed $\mu=0.5$ and different values of $\alpha$ (as explained by the legend). The parameters of the simulations are $L_{1}=40$ and $L_{2}=100$ and results shown here stem from averages over $10^7$ realizations. The lines are just guides for the eyes.}
\label{variance_simulation}
\end{figure}

Now, we compare the analytical expression (\ref{eq:variance}) with the estimate from numerical simulations looking for the values of $\alpha$ for which the two results converge. 
In Fig.~\ref{variance_simulation} we show numerical data for $\left\langle x^{2}(t)\right\rangle _{L}$, having posed $\mu=0.5$. In agreement with the previous numerical result, we see that for $\alpha\leq 0.5=\mu$ the power-law behavior (\ref{eq:variance}) is lost because the absorption can not be neglected. On the other hand, when $\alpha > \mu$ the asymptotic behavior allowing for absorption recovers the solution obtained for free diffusion. In fact, in this case $\rho(x,t)$ gets largely broad and absorption on the tails (corresponding to large values of $x$) is vanishing. 

Before concluding a few remarks are in order.\\
The simulations described so far have a cut-off in the jump length and in the size of the underlying chain because of the implementation on a computer (every data type has a maximum value) which transforms the motion into a truncated L\'evy flight on a finite chain. The latter is known to converge to a Gaussian statistics \cite{truncated} after suitably long times. Therefore, to study the L\'evy flight regime we have to consider only finite times. But this limitation is useful to get insights about finite-size systems, where the cut off is the natural constrain of the process. In fact, if the cut-off is large with respect to the interval of time, the walker actually does not perceive the finiteness of the system as if it effectively were in the thermodynamic limit. This allows us to understand that even in finite-size systems the case $\alpha>\mu$ with $\mu <1$ can lead to a dilation of trapping times. In fact, in the time spanned in our simulations, there is a finite probability of never being trapped when $\mu<\alpha$. Inevitably, this feature brings to a dilation of the mean trapping time also in finite structures. 


\section{ Numerical results for L\'evy walks and coupled continuous-time random walks} \label{sec:numerics2}

As explained in the introduction, animals often performs L\'evy walks. We now check numerically if the results obtained for L\'evy flights can also be applied to L\'evy walks which interact with the absorbing substrate only at the end of each jump. For instance, having in mind a bird exploring a given environment in the presence of a distribution $a(x)$ of targets, it can actually find the target while alighting on a given site between two consecutive flights.

From a naive point of view, this kind of process could be seen as a L\'evy flight with absorption, the only difference being the way time is running: for L\'evy flights the interval between two jumps (that is, between two interactions with the environment) is $1$, conversely for L\'evy walks the interval depends on the jump length.

\begin{figure}
\includegraphics[width=1.4\linewidth]{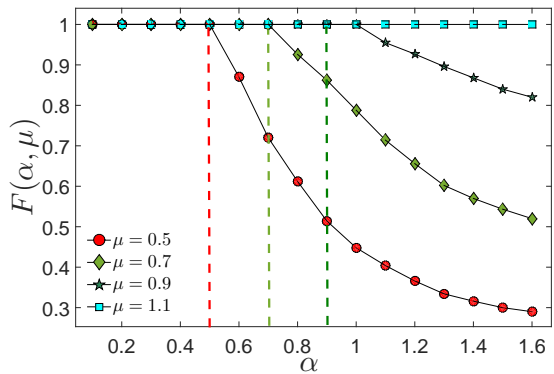}
\caption{(Color on line) The asymptotic absorption probability  $F(\alpha,\mu)$ found from Monte Carlo simulations for L\'evy walks. The cases analysed are the same as Fig.~\ref{F} and, again, absorption is certain for $\mu=1.1$ and for $\alpha\leq \mu$ with $\mu=0.5$, $\mu=0.7$ and $\mu=0.9$. The solid line are guides for the eye. The dashed vertical lines correspond to $\alpha=0.5$, $\alpha=0.7$, and $\alpha=0.9$. The number of replicas is $10^7$ and the asymptotic regime is reached typically around $t_f=10^6$. The errors on $F(\alpha,\mu) $ are of order $10^{-4}$.}
\label{Fabio}
\end{figure}

We therefore simulate a L\'evy walk in a one dimensional space by extracting $x$ from the probability distribution $p(x) =N(\mu)[|x|^{\mu+1}+1]^{-1}$ and by assuming the time taken to perform a jump $x$, to be proportional to $x$ itself.
After each jump the walker has a probability $a(x)$  of being absorbed, where $x$ is its current position.
Numerical simulations are performed following the same technique described in the previous section (first route) in such a way that we can derive the asymptotic probability of absorption $F(\alpha, \mu)$, shown in Fig.~\ref{Fabio}. Results are analogous to those obtained for L\'evy flights and evidence that also for L\'evy walks (with unitary velocity) the absorption is certain for $\mu\geq\alpha$.

We now check if the this property is true also for coupled continuous-time random walk.  
In particular, we introduce a coupled continuous-time random walk (e.g. see \cite{CCTRW}), whose stochastic behaviour converges to the one of the L\'evy flight we are interested in. More precisely, we first extract a waiting interval $\tau$ according to the power-law $\psi(\tau)\sim \tau^{-\beta}$ with $\beta>1$. The next jump will have a length $x$ depending on $\tau$, being drawn from the Gaussian distribution $\mathcal{N}(x; 0, \tau)$ with average $0$ and variance $\tau$. Thus, the longer the waiting time and the broader the distribution for the next jump length. The overall jump distribution turns out to be
\begin{eqnarray}
\nonumber
h(x) &\sim& \int \psi(\tau)\mathcal{N}(x; 0, \tau) \, d\tau =\int d \tau\,\tau^{-\beta}\frac{1}{\tau^{1/2}}\, e^{-\frac{x^{2}}{2\tau}}\\ 
&\sim& x^{1-2\beta}.
\label{eq: salti h(x)}
\end{eqnarray}
Therefore, the exponent $\beta$ characterising the waiting time $\tau$ rules the emerging L\'evy motion. By tuning $\beta$ we can recover different power law distributions for the jump lengths, being $\mu=2\beta-2$. The simulations are run long enough in order to reach the asymptotic regime, which is analysed as previously done for L\'evy flights and L\'evy walks, finding the same phase diagram and the same asymptotic absorption probability.

\section{Higher dimensions} \label{sec:higher}
Physical systems are often embedded in $2$ or $3$ dimensions. In these cases Eq.~\ref{eq:absorption_equation} takes the form
\begin{equation} \label{eq:absorption_d_dimensions}
\frac{\partial\rho(\vec{r},t)}{\partial t}=\nabla^{\mu} \rho(\vec{r},t)-a(\vec{r})\rho(\vec{r},t).
\end{equation} 
The asymptotic behaviour of the solution of the previous equation without absorption is known (e.g., see \cite{havlin}) and reads as
\begin{equation}\label{eq:radial}
\rho(r,t)\sim\frac{t}{r^{\mu+1}},
\end{equation}
which is analogous to Eq.~\ref{eq:normal}. This suggests that the analogy can be further extended to the absorbing case as long as the radial symmetry is retained, that is $a(\vec{r}) \equiv a(r)$. Indeed, Eq. \ref{eq:radial} shows that the jump length along the radial direction has a power low distribution as in the one-dimensional case. 
Therefore, considering the motion along the radial direction and a power law absorption $a(r)= K/ (r^{\alpha}+1)$, we realize that the condition provided by Eq.~\ref{eq:condition} for a non-null surviving probability is still valid.

\begin{figure}
\includegraphics[width=1.4\linewidth]{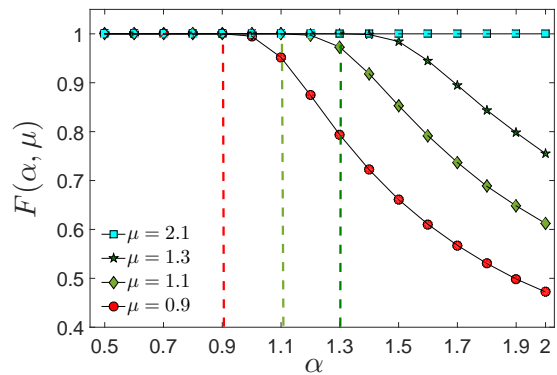}
\caption{(Color on line) The asymptotic absorption probability $F(\alpha,\mu)$ found from Monte Carlo simulations on a two-dimensional lattice. The data shown here are obtained according to the procedure explained in Sec.~\ref{sec:numerics}. Notice that the absorption is certain for $\mu=2.1$ and when $\alpha\leq \mu$ with $\mu=0.9$, $\mu=1.1$ and $\mu=1.3$. The solid line are guides for the eye. The dashed vertical lines correspond to $\alpha=0.9$, $\alpha=1.1$, and $\alpha=1.3$. The number of replicas is $10^7$ and the asymptotic regime is reached typically around $t_f=10^6$. The errors on $F(\alpha,\mu) $ are of order $10^{-4}$.}
\label{F2}
\end{figure}

On the other hand, Eq.~\ref{eq:condition2} holds only on a line. In fact, on a two-dimensional lattice, the condition for transience of a L\'evy flight process is $\mu<2$ \cite{Hughes}, in such a way that now avoiding absorption is feasible as long as
%
\begin{empheq}[left=\empheqlbrace]{align}
&\alpha>\mu, \label{eq:condition_2}\\
&\mu <2. \ \label{eq:condition2_2}
\end{empheq}
This picture is confirmed by the numerical results shown in Fig.~\ref{F2}.

\begin{figure}
\includegraphics[width=1.4\linewidth]{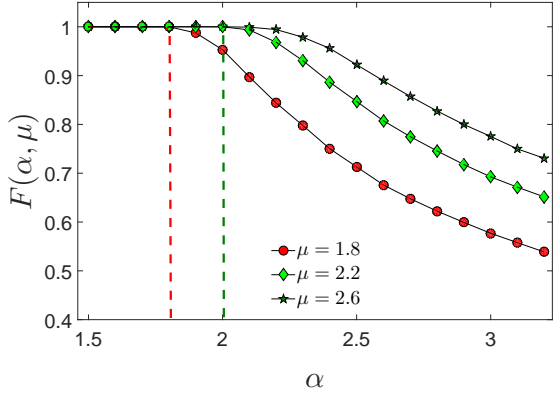}
\caption{(Color on line) The asymptotic absorption probability $F(\alpha,\mu)$ found from Monte Carlo simulations on a three-dimensional lattice. The data shown here are obtained according to the procedure explained in Sec.~\ref{sec:numerics}. Notice that the absorption is certain either with $\alpha\leq \mu$ when $\mu <2$ (e.g, with $\mu=1.8$) or with $\alpha <2$ when $\mu>2$ (e.g, with $\mu=2.2$ and $\mu=2.6$). The solid line are guides for the eye. The dotted vertical lines correspond to $\alpha=1.8$ and $\alpha=2$. The number of replicas is $10^7$ and the asymptotic regime is reached typically around $t_f=10^6$. The errors on $F(\alpha,\mu) $ are of order $10^{-4}$.}
\label{F3}
\end{figure}

Regarding the three dimensional case, we recall that this structure is transient, differently from the previous cases. Therefore, we expect that, for large values of $\alpha$, the absorption behaves like a finite size absorbing sphere in such a way that the walker has a finite probability not to be absorbed. In particular $a(x)$ has a finite average size when $\alpha>2$. Indeed, as successfully checked numerically (Fig.~\ref{F3}), the conditions to avoid the absorption are: 
%
%
\begin{equation}
\begin{dcases}
\alpha>\mu, \label{eq:condition_3}\\[1ex]
\mu <2.
\end{dcases}
\qquad \mbox{or} \qquad
\alpha>2.
\end{equation}

\begin{figure}
\includegraphics[width=1.4\linewidth]{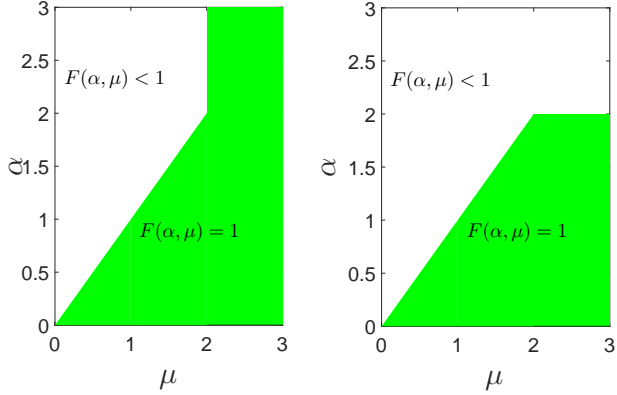} 
\caption{(Color on line) Phase diagram in the parameter space $(\alpha,\mu)$ for the asymptotic absorption probability of a Levy flight embedded in a two-dimensional (left panel) and in a three-dimensional (right panel) lattice. The white region corresponds to a finite surviving probability, that is $F(\alpha, \mu) <1$, while the colored region correspond to a certain trapping, that is $F(\alpha, \mu) =1$.}
\label{phase}
\end{figure}

Finally, the phase diagrams for the $2$ and $3$ dimension cases are shown in Fig.~\ref{phase} (left and right panel, respectively).

\section{ The velocity of the peaks} \label{sec:peaks}
In this section we provide some analytical arguments to corroborate the picture obtained in Sec.~\ref{sec:numerics}. 

\begin{figure}[tb]
\includegraphics[width=1.4\linewidth]{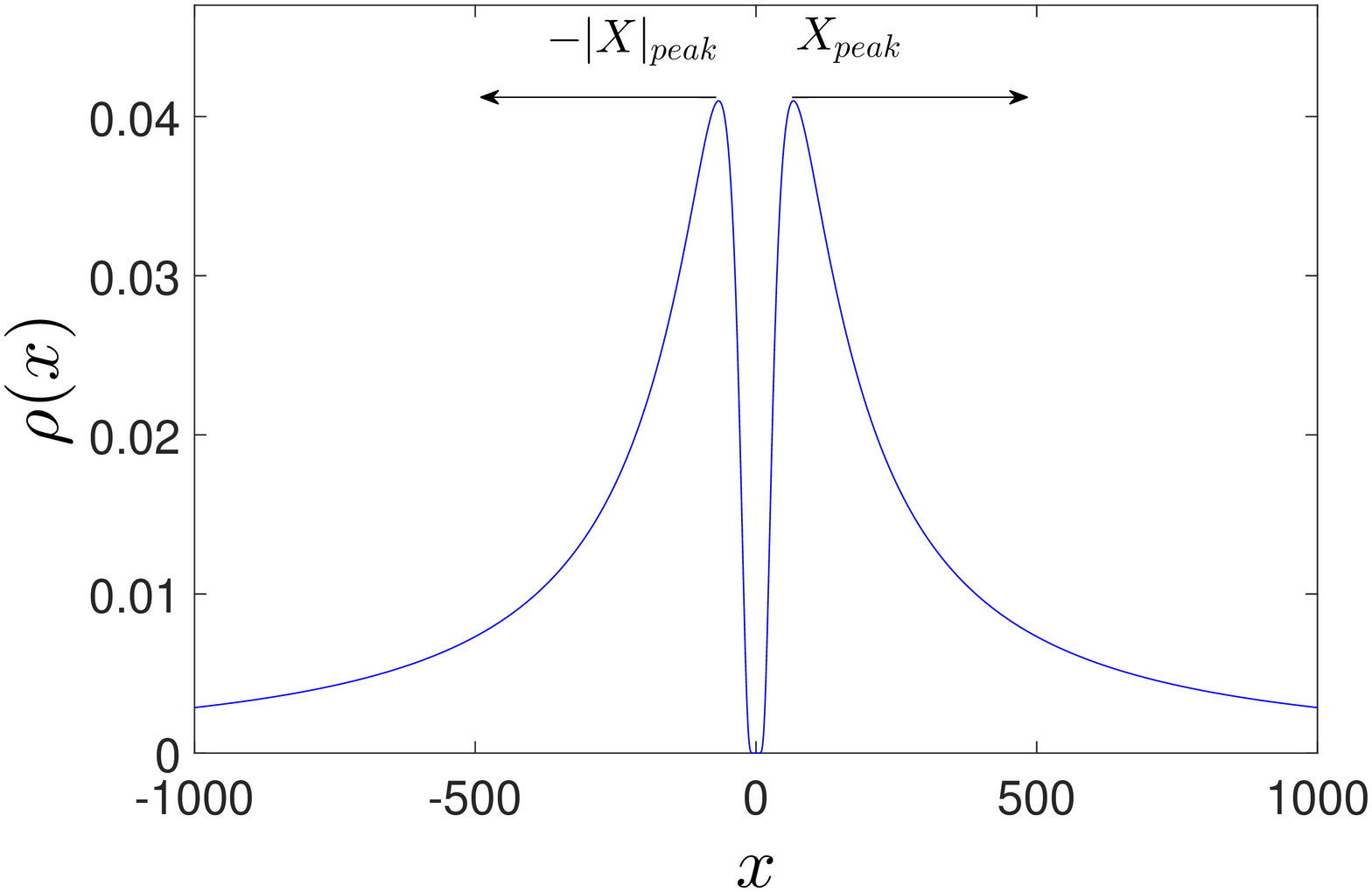}
\caption{(Color on line) The two expected peaks of the probability distribution $\rho(x,t)$. This plot is obtained from Eq.~\ref{eq:approximation} by fixing $\mu = 0.5$, $\alpha =1$ and $t=10^2$. \label{peak}}
\end{figure}

In the absence of trapping, the concentration $\rho(x,t)$, with initial condition $\rho(x,t) = \delta(x,0)$, evolves in time spreading out, yet remaining peaked at the origin \cite{Jespersen-PRE1999}. In the presence of the absorption term $a(x)$, the above description is changed as the absorption is certain in the origin.
Even in the neighbourhood of the origin the absorption is fast, because $a(x)$ is relatively large in this region. At the same time, for $|x|\to \infty$ the distribution $\rho(x,t)$ decreases (see Eq.~\ref{eq:asymptotic}). Then, after a short time, instead of a bell shaped curve, we expect a bimodal distribution with the two peaks propagating one on the right and one on the left (see Fig. \ref{peak}).

The distribution is always symmetric with respect to the origin, thus we will consider only the peak located at $x>0$ and try to estimate how its position, denoted with $X_{\textrm{peak}}(t)$, evolves in time. In particular, we distinguish two regions:
\begin{enumerate}
\item $0<x\ll X_{\textrm{peak}}(t)$, where the distribution is dominated by the absorption; in this region, a good approximation for $\rho(x,t)$ must decrease exponentially in time. In fact, the solution of Eq.~\ref{eq:absorption_equation} neglecting the diffusion term is $\rho(x,t)\sim\exp (- \frac{t}{x^{\alpha}+1})$;
\item $x\gg X_{\textrm{peak}}(t)$, where the distribution is dominated by diffusion, that is, $\rho(x,t)\sim t/(\delta +x^{\mu+1})$. Notice that $\delta$ is a positive number added in order to avoid the divergence of the power law in $x=0$ without changing the asymptotic behaviour for $x\to\infty$.
\end{enumerate}
Merging these remarks we can argue that a good approximation for the particle distribution may be:
\begin{equation}\label{eq:approximation}
\tilde{\rho}(x,t)\sim\exp\bigg(-\frac{t}{x^{\alpha}+1}\bigg)\,\frac{t}{\delta+ x^{\mu+1}}.
\end{equation}

In fact, Eq.~\ref{eq:approximation} recovers the two above-mentioned regimes for $x\sim0$ (where $\tilde{\rho}$ decreases exponentially in time) and $x\rightarrow\infty$, respectively. We also notice that, checking the normalisation,
\begin{equation}
\int_{0}^{\infty}\exp\bigg(-\frac{t}{x^{\alpha}+1}\bigg)\,\frac{t}{\delta+x^{\mu +1}}\, dx\sim t^{\frac{\alpha-\mu}{\alpha}},
\end{equation}
the approximation (\ref{eq:approximation}) can be used only for $\alpha<\mu$ to avoid divergences.


From Eq.~\ref{eq:approximation} we can find the position $X_{\textrm{peak}}$ of the peak by requiring
\begin{eqnarray}
\frac{\partial\tilde{\rho}}{\partial x} &\sim& \frac{\partial}{\partial x}\bigg[\exp\bigg(-\frac{t}{x^{\alpha}+1}\bigg)\,\frac{t}{\delta+x^{\mu +1}}\bigg]\\
&=& \frac{t}{\delta+x^{\mu +1}}\bigg[\frac{\alpha t x^{\alpha-1}}{(x^\alpha+1)^2}-\frac{x^\mu}{\delta+x^{\mu +1}} \bigg]=0.
\end{eqnarray} 
This leads to
\begin{equation}
\frac{\alpha t x^{\alpha-1}}{(x^\alpha+1)^2}=\frac{x^\mu}{\delta+x^{\mu +1}},
\end{equation}
and, solving for $t$, we get
\begin{equation}
 t =\frac{1}{\alpha}  \frac{x^\mu}{\delta+x^{\mu +1}} \frac{(x^\alpha+1)^2}{x^{\alpha-1}}.
\end{equation}
When $x$ and $t$ go to infinity:
\begin{equation} \label{eq:peak}
X_{\textrm{peak}}(t)\sim t^{\frac{1}{\alpha}}.
\end{equation} 

\begin{figure}[h]
\includegraphics[width=1.4\linewidth]{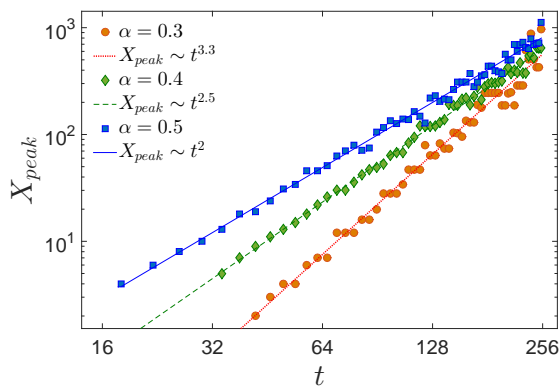}
\caption{(Color on line) The position of the peak with  $\alpha=0.3$, $\alpha=0.4$ and  $\alpha=0.5$. The graph is obtained by a Monte Carlo simulation and it confirms that $X_{\textrm{peak}}(t)\sim t^{\frac{1}{\alpha}}.$   \label{alpha04}}
\end{figure}

The scaling of Eq.~\ref{eq:peak} is confirmed by simulations (see Fig.~\ref{alpha04} where we show the cases  $\alpha=0.3$, $\alpha=0.4$ and  $\alpha=0.5$).
Therefore, in order to 
find the position of the peak it is sufficient to impose the behaviour of the distribution
near the origin and for $x\to \infty$ without a detailed information of the distribution 
in intermediate values of $x$. 


When $\alpha<\mu$ the velocity of the peak is larger than the velocity characterising the expansion of the asymptotic region, that is $x\sim t^{1/\mu}$ \cite{Jespersen-PRE1999}. Thus, in this case absorption prevails. On the other hand, when $\alpha>\mu$, the velocity of the peak is smaller than that of the asymptotic region, in such a way that the particle can escape without being absorbed. 

\section{conclusions} \label{sec:conclusions}

In this paper we investigated L\'evy flights and L\'evy walks in the presence of a power law distribution of traps (or targets). Different from the simple random walk, in this system trapping is, under proper conditions, not certain, no matter how long the process is. In fact, what matters is the interplay between the exponent $\mu$, ruling the distribution of step lengths, and the exponent $\alpha$, ruling the distribution of traps. 
Due to the great number of applications of the L\'evy model, the result can be exploited in several different contexts.
For instance, when $\mu$ is small (i.e. $\mu<1$ in one-dimensional lattices and $\mu<2$ in higher-dimensional lattices), as long as $\alpha > \mu$, a disease vector cannot prompt an epidemics.


Also, the differential equation studied in this paper may find interesting applications in finance. In fact, it is well known (see e.g., \cite{Mantegna}) that stock prices follow a stochastic behaviour, describable by L\'evy flights. Being $a(x)$ (not necessarily power-law) the agent inclination to sell a stock upon a variation $x$ of its original value, then the equation considered here could describe the temporal evolution of the probability that the stock is sold. To better mimic the evolution of the financial time series  it is possible to include the truncated L\'evy motion directly in the fractional equation \cite{frazionario_troncato}. Time correlations can also be included replacing the time derivative with a fractional derivative in time \cite{scalas}.

L\'evy walks and L\'evy flights have also been used to study models of foraging, usually considering a uniform distribution of traps (e.g., see \cite{Viswanathan-Nature1999}). The case of a uniform distribution of traps can be described in our model by setting $\alpha \to 0$. In this limit the particle is sure to be trapped for every value of $\mu$. For this reason, the power law trapping probability can be viewed as a generalization case which includes the uniform one. Further analyses should be done in order to understand what is the optimal strategy for locating sparse resources (as done in \cite{Viswanathan-Nature1999}) varying $\mu$ and $\alpha$. Indeed, the decreasing absorption probability could be the origin of the Brownian motion performed by some species, which do not follow L\'evy flights. Finally, this work may be a starting point to study the best strategy with a generic distribution of targets.

\appendix

\section{Fractional calculus and diffusion equation} \label{sec:fractional}

Fractional calculus is a natural extension of the differential and integral classical calculus. 
It has proved to be very useful in the study of diffusion processes; for instance, fractional differential equations are used to describe anomalous diffusion. In particular, when the jump distribution is $p(x)\sim|x|^{-1-\mu}$, the probability distribution $\rho(x,t)$ can be described by \cite{Metzler_Klafter,metzler2004restaurant}:

\begin{equation} 
\frac{\partial\rho(x,t)}{\partial t}=\frac{\partial^\mu \rho(x,t)}{\partial |x|^\mu},
\end{equation}
where $\mu\in\mathbb{R}$ and $\frac{\partial^\mu \rho(x,t)}{\partial |x|^\mu}$ is a derivative of fractional order $\mu$. We recall that, while classical derivation (for a generic function $f(x)$) is expressed through the symbols:
\begin{equation}
\frac{d^{n}f(x)}{dx^{n}};\,\ldots\,\frac{d^{2}f(x)}{dx^{2}};\,\frac{df(x)}{dx};\, f(x);
\end{equation} 
where the order of derivation is denoted by the index $n \in \mathbb{N}$, the fractional calculus allows extending the definition of the above mentioned operators to the case of real order, in fact the operators of arbitrary real order can be obtained with a sort of interpolation of the sequence of operators. 
Not like classical derivatives, there are several options for computing fractional derivatives (e.g., Riemann-Liouville, Caputo, Riesz, etc.).
Diffusion equations with fractional derivatives on space are usually written in terms of the symmetric Riesz derivative on space, whose operator is typically referred to as 
 $\nabla^\mu$. Now, this operator has a relatively simple behaviour under transformations, and, in fact, it is often defined in terms of its Fourier transform $\mathcal{F} [f]$ \cite{derivata_frazionaria,Jespersen-PRE1999} as
%

\begin{equation} \label{eq:fractional_derivative_definition}
\frac{\partial^\mu f(x)}{\partial |x|^\mu} = \mathcal{F}^{-1} [-|k|^\mu \mathcal{F} [ f(x)]].
\end{equation}

\section*{Acknowledgments}
\noindent
EA is grateful to GNFM and Spienza Universit\`a di Roma for financial support, and to Roberto Garra for interesting interactions. LC thanks Giovanni Cicuta for useful discussions. 

\bibliography{Omni}

\end{document}